\documentclass[12pt]{article}

\usepackage{bm}    
\usepackage{graphicx}   
\usepackage{sectsty}
\allsectionsfont{\normalsize}

\usepackage[fleqn]{amsmath}
\usepackage{amsmath}
\usepackage{amsfonts}   
\usepackage{amssymb}    
\usepackage{enumitem}		
\usepackage{braket}
\usepackage{ulem}
\usepackage{array}
\usepackage{perpage}
\DeclareMathAlphabet{\mathpzc}{OT1}{pzc}{m}{it}

\begin{document} 

\title{Reanalysis of Everett's relative-state formulation of quantum mechanics}  
\author{Jon Geist}
\date{2024 November}

\maketitle

\subsection*{Abstract}

\normalsize

\bigskip

\section*{} Everett's \lq\lq Relative State Formulation of Quantum Mechanics" (RSQM), which appeared in Reviews of Modern Physics, is based on his thesis \lq\lq The Theory of the Universal Wavefunction".  The most noteworthy property of these works is the claim by other authors that these works are the seminal contribution to Many Worlds theories of branching realities and the claim that practical laboratory experimental tests of RSQM are not possible.  This reports shows that Everett's two works describe a formulation of quantum mechanics that contradicts an overwhelming body of experimental physics.  Both works define a good observation of the value of a property of an object system initially in an eigenstate of a Hamiltonian operator as an interaction occurring in a larger, isolated system that also includes an observer system.  The RSQM definition of a good observation requires that the interaction between the object system and the observer system take place during a specified period of time and that the object system remain unchanged during that specified period of time. Consequently, a good observation cannot describe any physical observation that is carried out in a way that physically distinguishes the final state from the initial state.  Therefore, the set of good observations is a minuscule and completely negligible subset of physical observations. Nevertheless, both works assert that if the initial state of an arbitrary object system is a superposition of eigenfunctions then the final state of that system is given by a wavefunction with a very strange interpretation (strange wavefunction). This statement is demonstrably false.  The derivation of the strange wavefunction is based on the premise that all of the eigenfunctions in the initial superposition satisfy the stringent requirements imposed by the definition of a good observation.  Such wavefunctions are incapable of providing even a qualitative description of a vast majority of the measurements and observations reported in the literature of experimental physics. In other words, experimental tests of RSQM are not only possible, but have already been completed, and they universally reject RSQM as a meaningful model of the physical world. Of much less importance is that fact that neither work explicitly describes parallel worlds and HET explicitly states that the branching observer states occur within a single observer.  Most Many Worlds theories appear to be an attempt to extend or reinterpret Everett's interpretation of the strange wavefunction  in a more comprehensible way with little or no regard to the actual mathematical constraints imposed by the original formulation.  

\newpage

\section{\large{Introduction}}
In 1957, Everett published an article\cite{Everett01} with title \lq\lq Relative State Formulation of Quantum Mechanics" in Reviews of Modern Physics (HEA).  This article was based on his PhD thesis\cite{Everett02} \lq\lq The Theory of the Universal Wavefunction" (HET), which was not published until 1973. These works describe what he called a \lq \lq relative state" formulation of quantum mechanics (RSQM) in which a universal \lq\lq wavefunction that obeys a linear wave equation everywhere and at all times supplies a complete mathematical model for every isolated physical system without exception." 

The evolution of HET and HEA followed a tortured path.\cite{EverettBio} The thesis is 140 pages long and appears to be assembled from drafts of unpublished shorter reports\cite{EverettBio} that use somewhat different notation and terminology.  Furthermore, the notation is too concise to be clearly understood out of context. Also, some portions of HET and HEA appear to contradict other portions. This makes it difficult to determine exactly what is being asserted in these works. 

It is particularly vexing that Everett motivates both works by attacking the way that conventional quantum mechanics (CQM) treats change and then almost universally suppresses the time dependence in the representation of wavefunctions.   With this convention, time dependent wavefunctions are easily mistaken for time-independent wavefunctions in a casual reading of these works.  Both works define relative states of tensor products of wavefunctions but the notation fails to differentiate between factorable products, which represent independent systems, and unfactorable products, which represent interacting systems.

Nevertheless, when considered as a unit, the key ideas, equations, interpretations, and conclusions of HET and HEA are unambiguous as shown later in this report. However, as pointed out by Schlosshauer in his 2004 review paper, \lq\lq Decoherence, the measurement problem, and interpretations of quantum mechanics," which also appeared in Reviews of Modern Physics, \lq\lq Everett never clearly spelled out how his theory was supposed to work."\cite{Schlosshauer}. Strictly speaking, RSQM is not even a Many Worlds formulation because it describes a single observer in a type of superposition not allowed in CQM instead of multiple observers in different parallel worlds.  

\subsection{Summary description and critique of RSQM}

Both HET and HEA define a good observation of the value of a property of a system initially in an eigenstate of a Hamiltonian operator as an interaction occurring in an isolated total system.  Both works partition the total system into a composite system that consists of an object system (system under study) and an observer system. The total system is sometimes referred to as the system, sometimes as the composite system, and rarely as the total system.  Unfortunately, the object  system is also  referred to as the system in one portion of HET.  For clarity, this report adds words in square brackets to quotations from HET and HEA to clarify nomenclature that might be ambiguous out of its larger context.   

Up to a point RSQM cannot be proven false because it is based not only on postulates but also on definitions, which are neither true nor false.  Instead of truth values, definitions are judged by the fraction of the pertinent subject matter to which they apply.  However, Everett (apparently unintentionally) falsifies RSQM on page 458 of HEA (and in slightly different notation and wording on page 66 of HET), where he asserts,
\begin{quote}
\textit{From the definition of a good observation we first
deduce the result of an observation upon a system which is \textbf{not} in an eigenstate of the observation. We know from our definition that the interaction transforms states
$\phi_i \psi^O[\dots]$ into states $\phi_i \psi^O[\dots\alpha_i]$. Consequently these solutions of the wave equation can be superposed to give the final state for the case of an arbitrary initial [object] system state. Thus if the initial [object] system state is not an eigenstate, but a general state $\sum_i a_i \phi_i$, the final total state will have the form:
\begin{equation}
\tag{HEA 12} 
\label{S'+O'}
\psi^{S+O'} = \sum_i a_i \phi_i \psi^O[\dots \alpha_i].
\end{equation}
This superposition principle continues to apply in the presence of further systems which do not interact during the measurement.}
\end{quote} 

This assertion is logically indefensible.  Equation HEA 12 does not describe the final state for an arbitrary initial-state superposition of eigenfunctions. It only describes the final state for an initial-state superposition of eigenfunctions that satisfy the definition of a good observation.  

However, the definition of a good observation imposes a stringent requirement on an object system that is in an eigenstate:  [HEA page 458 and HET page 64] \lq\lq That is, we require that the [object] system state, if it is an eigenstate, shall be unchanged, ..." during the entire observation.  This requirement rules out an overwhelming majority of the physical observations described in the physics and chemistry literature since these observations consist of sequences of transitions between different initial and final states with each final state serving as the initial state of the next transition.  

On the face of it, the assertion about Eq. HEA 12 appears preposterous.  However, within the context of HET, it is not such a gigantic extrapolation.  In HET, prior to the definition of a good observation, Everett defined an RSQM measurement in almost the same way as an RSQM good observation.  Most importantly, that definition imposed the same stringent requirement on an RSQM measurement as that imposed on an RSQM good observation.  HET then presented a one-dimensional example of a measurement of the value an arbitrary property of an arbitrary object system by an arbitrary observer system to illustrate the generality and power of the RSQM definition of a measurement.  

Unfortunately, the example contained a fatal mathematical error, which restricted the RSQM measurement to object systems represented by wavefunctions that are constant in time and space.  This rules out most if not all of experimental physics and chemistry.  However, without knowledge of this error and with false confidence in the generality and power of the definition of an RSQM observation, extrapolation of the results derived for an RSQM observation to all observations does not appear so large a stretch.  This of course does not excuse what is a major blunder, but it does perhaps explain it. 

Later in HET, Everett interpreted Eq. HEA 12 as describing a single observer that simultaneously observes each eigenstate in a superposition as the only state in the superposition: 
\begin{quote}
\textit{Whereas before the observation we had a single observer state afterwards there were a number of different
states for the observer, all occurring in a superposition. Each of these separate states is a state for an observer, so that we can speak of the different observers described by the different states. On the other hand, the same physical system is involved, and from this viewpoint it is the same observer, which is in different states for different elements of the superposition (i.e., has had different experiences in the separate elements of the superposition). In this situation we shall use the singular when we wish to emphasize that a single physical system is involved, and the plural when we wish to emphasize the different experiences for the separate elements of the superposition. (e.g., "The observer performs an observation of the quantity A, after which each of the observers of the resulting superposition has perceived an eigenvalue.")} 
\end{quote}

Of course, by this point, RSQM is already debunked as science fiction or a theory that describes a negligible fraction of all physically meaningful wavefunctions included in any valid theory of a universal wavefunction. Apparently unaware of this problem, but aware that his interpretation of Eq. HEA 12 was likely to encounter objections, Everett provided an explanation to counter such objections. 

\subsection{RSQM's last gasp} 

Everett attempted to address the difficulty of conceptualizing a single observer that simultaneously observes each eigenstate in a superposition as the only state in the superposition.  He attributed this difficulty to a bias created by the influence of the concepts of CQM on the subjective nature of perception.  Indeed HET mentions \lq\lq subjective" 18 times in this connection.  This argument may have some merit in discussions about how to think about Eq. HEA 12, but it fails completely when applied to justify the idea that object systems do not change during measurements of their properties.  

The words \lq\lq same", \lq\lq different", and \lq\lq change" are at the heart of the human experience and have meanings that are independent of physical theories. Experiments are attempts to characterize our experience in such a way that we can predict future experiences with some level of uncertainty based on our past experiences. Only a sophist with an extreme relativist agenda would claim that the state of an object system comprising a tritium atom was the same following the radioactive decay of that atom into helium-3, or that the state of the Ca$^{++}$ ions in a sample of CaCl$_2$ dissolved in water was identical to the state of the Ca atoms in the Ca(OH)$_2$ precipitate created by mixing CaCl$_2$ solution with a solution of NaOH.  

In essence, RSQM was apparently unintentionally derived to describe a subset of unknown size of all phenomena and used to draw conclusions about that subset of phenomena.  These conclusions were then asserted to apply to all phenomena. Since the subset of phenomena that RSQM can describe turns out to be negligible, RSQM is a failed program.  

In contrast, the mathematical formulation of CQM was derived to describe the subset all of the experimental evidence available about atomic spectroscopy and radiochemistry at its inception.\cite{Kirby}  Bohr and others came up with an interpretation of CQM based on the Bohr-Sommerfeld picture.  In some quarters  this interpretation outlasted its usefulness and plausibility.  Others corrected and extended CQM to encompass new phenomena as well as phenomena previously thought to fall outside its scope. The improvement of the accuracy and scope of CQM was mostly characterized by the constant interaction of theory and experiment: theory suggesting new experiments and challenging some experimental results, and experimental results honing and pruning theory. 

In retrospect, RSQM was an audacious attempt to simultaneously address a number of objections to CQM, many of which were already discredited, by an approach which postpones comparison with experiment until most of its development was completed.  This attempt failed for a number of reasons associated with this approach.   
 
\subsection{Digression: Can RSQM be salvaged}

At this point, it is reasonable to wonder whether RSQM can be salvaged by eliminating the stringent requirement imposed by the definition of a good observation without morphing into CQM.  This question is well beyond the scope of this report because it would require a similarly in-depth review (from the perspective of this report) of all of the variety of elaborations and variations of RSQM published over the years.  

Personally, I don't think RSQM can be salvaged because the portions of RSQM that are based on stringent requirements have to be replaced if it is to describe all of experimental physics and chemistry.  For instance, CQM describes experimentally verified transitions from initial states to different final states of vapor-phase atoms and ions in terms of absorption or emission of one or more photons.  These ideas, which are central to experimental atomic physics, are based on spectroscopy, the photoelectric effect, etc. These same experimentally inspired ideas are completely alien to RSQM. 

For instance, neither publication mentions photons even once. Correlations, which replace transitions in RSQM, are mentioned 125 times in HET:  Transitions are mentioned only 10 times in HET: 3 times in connection with information theory, 2 times to qualitatively describe transitions from relative eigenstates of one operator into relative eigenstates of a different operator, and 5 times to deny the validity of transitions from initial states to final states as fundamental process in quantum physics.  

Correlations are mentioned 13 times in HEA. Transitions are mentioned only 6 times in HEA, mostly to deny their existence: 4 times to explain that there is no "transition from possible to actual" in RSQM, one time to claim that RSQM calculates the probability for each initial observation by the use of the usual transition probabilities (even though that calculation requires the final state of a system in an eigenstate be different from the initial state of the system, which contradicts the RSQM definition of a good observation), and one time to claim that CQM cannot describe the entire universe because \lq\lq There is nothing outside it to produce transitions from one state to another."    

\subsection{After publication}

Initially, HEA was not well received by the physics community.\cite{Byrne01}  Ballentine's\cite{Ballentine} 1970 defense of the statistical interpretation of quantum mechanics, which also appeared in Reviews of Modern Physics, never cites or even mentions RSQM, HEA, HET, or Everett while devoting over four pages to the measurement problem.  

Ironically, in the same year that Ballentine's paper was published, Bryce DeWitt, who initially rejected the reality of observer branching\cite{EverettBio}, claimed in a Physics Today article\cite{DeWitt1} (also without spelling out how RSQM actually worked) that RSQM resolved Schroedinger's cat paradox (as translated into English in \cite{Trimmer}). In fact, it was DeWitt who popularized the name \lq\lq many worlds".  This term is more charismatic than \lq\lq relative state formalism" and emphasizes the most striking difference between De Witt's interpretation of RSQM and conventional quantum mechanics (CQM), but never appears in \cite{Everett01} or \cite{Everett02}.  

DeWitt's article implicitly rejected Everett's interpretation of HEA 12 as describing a single observer.  Instead DeWitt assigned each term in Eq. HEA 12 to a different observer in a different universe (world) and explicitly recognized that his interpretation of RSQM requires the splitting of one universe into multiple almost identical copies at each splitting event, and also implicitly recognized that the number of observers of a good interaction must be an integer which sets the minimum number of universes created following each observation equal to the reciprocal of the smallest non-zero transition probability associated with the observation. 

Clearly DeWitt believed that reality branching described all interactions, not just RSQM good observations and RSQM measurements, because he explicitly stated \cite{DeWitt1} 
\begin{quote}
\textit{Moreover, every quantum transition taking place on every star, in every galaxy, in every remote corner of the universe is splitting our local world on earth into myriad of copies of itself... I still recall vividly the shock I experienced on first encountering this multiworld concept. The idea of $10^{100+}$ slightly imperfect copies of oneself all constantly splitting into further copies, which ultimately become unrecognizable, is not easy to reconcile with common sense.}
\end{quote} 

DeWitt's article triggered a growth of interest in Everett's ideas through another tortuous sequence of events \cite{EverettBio}.  DeWitt also stated without justification \cite{DeWitt1}, 
\begin{quote}
\textit{Clearly the EWG\footnote{EWG is DeWitt's acronym for the what he calls a meta-theorem, which states \lq\lq the mathematical formalism of quantum mechanics as it stands without adding anything to it is all that is needed to solve the problem of describing observations, which was first given in 1957 by Hugh Everett with the encouragement of John Wheeler and has been subsequently elaborated by R. Neil Graham.\cite{DeWitt1}"} view of quantum mechanics leads to experimental predictions identical with those of the Copenhagen view.  This, of course, is its major weakness. Like the original Bohm theory$^{\;6}$ it can never receive operational support in the laboratory. No experiment can reveal the existence of the \lq\lq other worlds" in a superposition like that in equations 5 and 6.} 
\end{quote}  

This statement apparently bolstered belief in the superficially plausible idea that a search for laboratory tests of RSQM would be a waste of time. Currently, many physicists and philosophers treat reality branching, including that in DeWitt's extension of RSQM to the cat paradox, as if it is an untestable but perhaps inevitable consequence of quantum mechanics.\cite{Kent} In fact, some even believe that RSQM is the only accurate version of quantum mechanics.      

To complete this summary, this report asserts that RSQM is demonstrably flawed as summarized above and carefully critiqued below. If this conclusion is independently validated, future technical and popular references to it should assert this fact rather than continue to characterize it as a possible description of reality.   

\section{Detailed critique of RSQM}

\subsection{General comments}

As described above, the evolution of HET and HEA (HE) followed a tortured path.\cite{EverettBio} The thesis is 140 pages long and appears to be assembled from drafts of unpublished shorter reports\cite{EverettBio} that use somewhat different notation and terminology.  This sometimes makes it difficult to determine exactly what is being asserted in different portions of HET and HEA. 

The explicit time dependence of the mathematical quantities is suppressed in most symbols and equations.  This is surprising in publications about change. It is possible that this concise notation led to the grossly misleading error in the example of a measurement.  

The terms, \lq\lq measurement" and \lq\lq good observation" are defined in HE. The term \lq\lq observation"is explicitly defined as a \lq\lq measurement" at least once in RSQM: 
\begin{quote}
HEA 457 and HET 64: \textit{Such a machine will then be capable of performing a sequence of observations (measurements), \dots} 
\end{quote}
The phrases \lq\lq measurement or observation" and \lq\lq observation or measurement", which could imply that they are synonyms, but could also imply that they have different meanings, are also used.  HET and HEA appear more internally coherent and consistent with each other if \lq\lq measurement" and \lq\lq observation" are synonyms even though RSQM good observations occur during a \lq\lq specific period of time" and RSQM measurements are defined in the limit of infinite time.

The terms \lq\lq apparatus" and \lq\lq observer" are explicitly defined as synonyms at least once in RSQM: 
\begin{quote}
HEA 101: \textit{An approximate measurement consists of an interaction, for a finite time, which only imperfectly correlates the apparatus (or observer) with the object-system.}
\end{quote}       

\subsection{HE begins by criticizing how CQM treats change}

\begin{quote}
 HET 3 and HEA 454: \textit{We begin, as a way of entering our subject, by characterizing a particular interpretation of quantum theory which, although not representative of the more careful formulations of some writers, is the most common form encountered in textbooks and university lectures on the subject. \dots \newline
\mbox{\;\;\;\;}Thus there are two fundamentally different ways in which the state function can change: \newline
\mbox{\;\;\;\;}\textbf{Process 1:} The discontinuous change brought about by the observation of a quantity with eigenstates $\phi_1, \phi_2, \dots$, in which the state $\psi$ will be changed to the state $\phi_1$ with probability $(\phi_1, \phi_j)^2$. \newline
\mbox{\;\;\;\;}\textbf{Process 2:} The continuous, deterministic change of state of the (isolated) system with time according to a wave equation $\partial \psi/\partial t = U \psi$, where U is a linear operator.}  
\end{quote}
HE then argues that the inclusion of both discontinuous and continuous descriptions of change produces paradoxes in CQM in the case of multiple observers.
\begin{quote}
\textit{The question of the consistency of the scheme arises if one contemplates
regarding the observer and his object-system as a single (composite)
physical system. Indeed, the situation becomes quite paradoxical if we
allow for the existence of more than one observer. Let us consider the
case of one observer A, who is performing measurements upon a system S,
the totality (A + S) in turn forming the object-system for another observer,
B.
If we are to deny the possibility of B's use of a quantum mechanical
description (wave function obeying wave equation) for A + S, then we
must be supplied with some alternative description for systems which contain
observers (or measuring apparatus). Furthermore, we would have to
have a criterion for telling precisely what type of systems would have the
preferred positions of "measuring apparatus" or "observer" and be subject
to the alternate description. Such a criterion is probably not capable
of rigorous formulation.
On the other hand, if we do allow B to give a quantum description to
A + S, by assigning a state function $\psi^{A+S}$, then, so long as B does not
interact with A + S, its state changes causally according to Process 2,
even though A may be performing measurements upon S. From B's point
of view, nothing resembling Process 1 can occur (there are no discontinuities),
and the question of the validity of A's use of Process 1 is raised.
That is, apparently either A is incorrect in assuming Process 1, with its
probabilistic implications, to apply to his measurements, or else B's state
function, with its purely causal character, is an inadequate description of
what is happening to A + S.}
\end{quote}

This argument does not survive close scrutiny. First, the discontinuity in Process 1 is crucial only to the Bohr-Sommerfeld theory of quantum mechanics. It is completely consistent with CQM to consider Process 1 as an extremely useful approximation to Process 2 for abrupt (but continuous) transitions that cannot or need not be resolved or adequately modeled in an application of interest. 

In this case, the discontinuous jump conveniently replaces the unresolved processes that occur during the transition.  This approximation is most useful when the initial and final states of the process are known, in which case the average transition rate can be calculated with Fermi's golden rule or one of its extensions or elaborations and reasonable approximations to the initial state Hamiltonian and the interaction Hamiltonian.  

\subsection{HET introduces and initiates the RSQM program}

\begin{quote}
HEA page 455: \textit{This paper proposes to regard pure wave mechanics (Process 2 only) as a complete theory. It postulates that a wave function that obeys a linear wave equation everywhere and at all times supplies a complete mathematical model for every isolated physical system without exception. It further postulates that every system
that is subject to external observation can be regarded
as part of a larger isolated system. \newline
\mbox{\;\;\;\;} The wave function is taken as the basic physical entity with no a priori interpretation. Interpretation only comes after an investigation of the logical structure of the theory. Here as always the theory itself sets the framework for its interpretation.}
\end{quote}
HET implemented this proposal in terms of an object system and an observer system by omitting the special postulates in CQM that deal with observation and replaced them with a new RSQM theory of observation based on using correlations to describe interactions between the object and observer systems.  This is very different from the CQM approach based on sequences of transitions from one system state to another system state\footnote{where state and system are flexibly defined: The tritium and helium-3 nuclei are different nuclear systems or different states of the three-nucleon nuclear system depending on point of view.}. The implementation started by optimizing information theory and what remained of CQM after omitting the special postulates in CQM that deal with observation.  The result was the foundation for developing the new RSQM theory of observation. 

There are two potential problems with this approach. HET eventually acknowledges the first of these problems on page 90:
\begin{quote}
\textit{In Chapter III and IV we discussed abstract measuring processes, which were considered to be simply a direct coupling between two systems, the object-system and the apparatus (or observer). There is, however, in actuality a whole chain of intervening systems linking a microscopic system to a macroscopic observer. Each link in the chain of intervening systems becomes correlated to its predecessor, so that the result is an amplification of effects from the microscopic object-system to a
macroscopic apparatus, and then to the observer.}
\end{quote}

The potential problem with lumping all of these links together into a single observer system is that assumptions which appear plausible for all lumped systems might be demonstrably false for some linked systems. The other potential problem is the implicit assumption that correlation of the properties of two physical systems implies that the two systems are directly interacting. 

For instance, consider one detector measuring the radiation from a source that is transmitted through a beam splitter and a second detector that is measuring the radiation reflected by the same beam splitter.  These detectors do not actually interact with each other even though their outputs are correlated.  Instead, both detectors interact with two different radiation fields that are correlated.   

Since it is possible to set up a third radiation field that interacts with only one of the detectors, the two detectors actually evolve independently of each other even though their outputs are correlated by their interaction with a subset of fields that are correlated.  

In any case, Everett did encounter a problem with this approach: 
\begin{quote}
HET 53: \textit{Nearly every interaction between systems produces some correlation however. Suppose that at some instant a pair of systems are independent, so that the composite system state function is a product of subsystem states $(\psi^{S} = \psi^{S1} \psi^{S2})$. Then this condition obviously holds only instantaneously if the systems are interacting --- the independence is immediately destroyed and the systems become correlated. We could, then, take the
position that the two interacting systems are continually "measuring" one another, if we wished. At each instant $t$ we could put the composite system into canonical representation, and choose a pair of operators $\breve{A}$ in $S_1$ is measured by $\breve{B}$ in $S_2$ which define this representation. We might then reasonably assert that the quantity $\breve{A}$ in $S_1$ is measured by $\breve{B}$ in  $S_2$ (or vice-versa), since there is a one-one correspondence between their values. \newline
\mbox{\;\;\;\;}Such a viewpoint, however, does not correspond closely with our intuitive idea of what constitutes measurement.}
\end{quote}  

In fact, the asymmetry comes from the details of the transfer of information between each link in the chain from microscopic to macroscopic. For instance, when an atom in an excited state emits a photon, both the internal state of the atom and the change in the occupation number of the pertinent mode of the photon field temporarily record this event. However, the internal state of the atom can transition to a lower state while the state of the field remains entangled with the center of mass of the atom until the emitted photon is annihilated (absorbed), etc. 

This fact is how we know that radioactive atoms are not entangled with their decay products until shortly before the decay event.  This is most obvious in the case of long-lived beta emitters like tritium.  If each tritium atom became entangled with its $^3$He$^+$ decay product from its formation until it emitted a beta particle, there would be plenty of time to measure the physical and chemical properties of the entangled product. For instance, it would be possible to separate a mixture of tritium, the entangled product, and helium-3 by fractional distillation.  However no intermediate fraction has ever been reported. 

Unfortunately, there is no way to directly observer the internal state of an atom.  The only way is to infer its initial state from the absorption and/or emission of photons during which its state changes. Furthermore, macroscopic systems have emergent properties such as easily measured voltages that are very useful for encapsulating information in compact and convenient formats.   In other words, the asymmetry is related to what is physically measurable and how to record and store it compactly and practically.  

\subsection{HET introduces an unnecessary asymmetry into RSQM}

Apparently, Everett was not aware of this situation and decided that it was necessary to introduce a fundamental asymmetry into any interaction that involved an object system and an observer system independent of what is and is not physically measurable. This is really surprising because his stated goal was quite the opposite:  
\begin{quote}
HET 53: \textit{From our point of view there is no fundamental distinction between \lq\lq measuring apparata[sic]" and other physical systems.}
\end{quote}
Yet he asserts:
\begin{quote}
HET page 65 and HEA 458: \textit{That is, we require that the [object-]system state, if it is an eigenstate, shall be unchanged, and (2) that the observer state shall change so as to describe an observer that is \lq\lq aware" of which eigenfunction it is; that is, some property is recorded in the memory of the observer which characterizes $\phi_i$, such as the eigenvalue.}
\end{quote}

The problem here is that properties of mathematical correlations of values of properties do not control interactions. It is the other way around.  The properties of interactions determine the mathematical correlations of values of properties.  By failing to recognize that an object system does not have to remain in its initial state until the record of that state is transferred through any number of intermediate links to a (semi)permanent record of that state, Everett restricted the scope of RSQM. But this restriction is just the beginning of the problems with RSQM. 

\subsection{HET extends unnecessary asymmetry to wavefunctions}

HET introduces the first restriction to the RSQM definition of a measurement on pages 53-55. In the last paragraph of a discussion about how to add asymmetry to correlations meeting the requirements of RSQM measurements prior to (rather than following) the RSQM definition of a measurement, HET highlights what turns out to be a fatal consequence of this requirement:   
\begin{quote}
\textit{The restriction that $H$ shall not decrease the marginal information of [the value] $A$ has the \textbf{interesting consequence} that the eigenstates of [the operator with eigenvalue] $A$ will not be disturbed, i.e., initial states of the form $\psi^S_t = \phi \eta_0$ where $\phi$ is an eigenfunction of $A$, must be transformed after any time interval into states of the form $\psi^S_t = \phi \eta_t$, since otherwise the marginal information of $A$, which was initially perfect, would be decreased. This condition, in turn, is connected with the repeatability of measurements, as we shall subsequently see, and could alternately have been chosen as the condition for measurement. }
\end{quote}
 
The meaning of this paragraph may not be immediately clear even after reading the prior discussion in HET because the notation is so concise.  Let the wavefunction $\phi(t, \vec{r})$ describe the behavior of the object system during the measurement. What exactly does $\phi$ mean?  Is $\phi$ independent of time, that is to say $\phi =  \phi(t_0, \vec{r})$ for some time $t_0$, or is a less concise notation needed to clarify the meaning. 

Using a notation chosen to be similar to the concise notation used HET, let 
\begin{equation}
\label{initial}
\psi^{S+O}(t) = \phi_i(t) \eta(t) 
\end{equation}
for $t \ge t_0$
represent the initial state of the system $S$, and let   
\begin{equation}
\label{final}
\psi^{S'+O'}(t) = \phi_j(t) \xi(t) 
\end{equation}
for $t \ge t_0$
describe the continuous time dependent evolution of the system $S$ to its final state.  According to the paragraph under consideration, the wavefunction $\eta(t)$ evolves from $\eta(t_0)$ to $\xi(t) \ne \eta(t)$ for $t \ge t_0$.    
This still leaves $\phi_j$ undetermined.  

One possibility is that $\phi_j(t) = \phi_i(t_0)$. This option describes a zero-energy state in a quantum field theory. The mass and momentum of an object system in this state are both identically zero.  Such states are rare, esoteric, and not at all typical of the non-zero energy states that are the heart and soul of experimental quantum mechanics and the bread and butter of theoretical quantum mechanics. 

Zero-energy states that just convert open intervals into closed intervals form a completely negligible subset of the continuum spectrum of eigen-distributions.  They are completely negligible in the sense that the value of any physically meaningful integral over a continuum of eigen-distributions is independent of whether or not this subset is included when the integral is evaluated. 

If this is the RSQM definition of a measurement, then RSQM measurements are completely irrelevant.  The only other plausible option is that $\phi_j(t) = \phi_i(t)$ for all $t$.  This is equally problematic because it rules out all experiments for which the final state of a system following an interaction is different from the initial state of the system at the beginning of the interaction where not only $\xi(t) \ne \eta(t)$ but also $\phi_j(t) \ne \phi_i(t)$ for some times $t > t_0$ in Eq. \ref{final}. 

At this point, RSQM is completely useless unless or until someone describes a wavefunction that (without any change in the form of the wavefunction and without any change in the quantum numbers in the wavefunction) can describe two different physical systems or two different states of the same physical system. (Here we start to use HE to refer to both HEA and HET as a singular body of work.)    

\subsection{HE presents a fatally flawed example of an RSQM measurement process} 

Immediately following the RSQM definition of a measurement and apparently unaware of the limitations of this definition, HET presents an example of a measurement process governed by a Hamiltonian that appears at first glance to justify the HE definition of a measurement. HEA presents the same example on page 456. 

It is very important to understand that that this example does not justify the HE definition of a measurement because it contains a fatal mathematical error.  In fact, this example suggests that the RSQM definition of a measurement is so exclusive as to render it irrelevant to experimental and theoretical physics in general, and even more so to any theory of a universal wavefunction.

\subsubsection{The flawed example} 

On page 456 of HET and page 56 of HEA (coincidence no doubt), HE states
\begin{quote} \textit{
At this point we consider a simple example, due to von Neumann, which serves as a model of a measurement process. Discussion of this example prepares the ground for the analysis of \lq\lq observation." We start with a system of only one coordinate, $q$ (such as position of a particle), and an apparatus of one coordinate $r$ (for example the position of a meter needle). Further suppose that they are initially independent, so that the combined wave function is $\psi_0^{S+A}=\phi(q)\eta(r)$ where $\phi(q)$ is the initial system wave function, and $\eta(r)$ is the initial apparatus function. The Hamiltonian is such that the two systems do not interact except during the interval $t=0$ to $t= T$, during which time the total Hamiltonian consists only of a simple interaction\footnote{The HEA and HET versions of this paragraph differ somewhat.  Therefore, the more recent version is quoted here.} , 
\begin{equation}
\tag{HEA 4} \label{HEA 4}
H_I = -i\hbar q(\partial/\partial r).
\end{equation}
Then the state
\begin{equation}
\tag{HEA 5} 
\psi_t^{S+A}(q,r) =  \phi(q)\eta(r-qt)
\end{equation}
is a solution of the Schroedinger equation,
\begin{equation}
\tag{HEA 6} 
i\hbar (\partial \psi_t^{S+A}/\partial t) = H_I \psi_t^{S+A},
\end{equation}
for the specified initial conditions at time t= 0.}
\end{quote} 

\subsubsection{The error in the example}

This example, which employs an overly concise notation, is presented as if it applies to any one-dimensional eigenfunction $\phi(q,t)$ and any one-dimensional apparatus system $\eta(r,t)$.  Therefore it is very important to understand that Eqs. HEA 5 and HEA 6 are not true in general. In fact they are true only for exceedingly few and very special quantum states.  

In general,
\begin{equation}
i\hbar \partial \psi_t^{S+A}/\partial t = H_I \psi_t^{S+A} + i \hbar\left(\frac{\partial \phi(q,t)}{\partial t}\eta(r-qt)\right).
\end{equation}
Therefore, Eqs. HEA 5 and HEA 6 can be true only if $\eta(r-qt) = 0$ during the interaction (which is not possible for any CQM wavefunction), or $\partial \phi(q,t)/\partial t = 0$ for $0 \le t \le T$.  

It is ironic, that in the latter case, $\phi(q,t)$ is the zero-energy state in a quantum field theory that was rejected as a possible interpretation of the scope of the RSQM definition of a measurement because it renders RSQM completely irrelevant to physics.  

Apparently, without knowledge of the limitations placed on $\phi$ by Eq. HEA 6, HET completed this section of the thesis with a discussion of how to use the results of the example measurement.  Since these results depend upon the assumption that $\partial \phi(q,t)/\partial t = 0$, they are useless unless independently derived from more widely applicable postulates.   

However, this fallacious example of an RSQM measurement does not disprove RSQM because a definition cannot be disproved.  Instead, this definition must be judged by its scope, that is to say the size, range, and depth of the body of experimental physics that it can describe compared to that of CQM. The short answer is its scope falls somewhere between a negligible fraction of experimental physics and none at all.   

This fallacious example of a general RSQM measurement was based on a fantasy: the existence of magic-like Hamiltonians that can interact with an object system in such a way as to extract information from that system and encode that information in an apparatus system with no change whatsoever to the state of the object system. 

At this point, RSQM is already dead.  Even so, the following material is presented for anyone who is curious about the continued development of RSQM to it preposterous conclusions.  Apparently, not only unaware of its fatal limits, but with an unjustified confidence in the generality of the RSQM definition of a measurement based on a mathematical error in an example of the RSQM measurement process, HET proceeds to develop an abstract theory of observation in $\textit{Chapter IV. Observation}$ as the culmination of the RSQM program. 

\subsection{RSQM definition of an observer system}

The definition of an observer state is given pages 64 and 65 of HET and in almost identical form on page 458 of HEA):  
\begin{quote}
\textit{When dealing quantum mechanically with a system representing an observer we shall ascribe a state function, $\psi^O$, to it. When the state $\psi^O$ describes an observer whose memory contains representations of the events $A,B,\dots,C$ we shall denote this fact by appending the memory sequence in brackets as a subscript, writing\footnote{The HEA and HET versions of this paragraph are identical except that HET has no equation numbers, so HEA is used as the source for this material.}:
\begin{equation}
\tag{HEA 9} \label{S+O}
\phi^O[A,B, ... ,C]. 
\end{equation}
The symbols $A,B, ... ,C$, which we shall assume to be ordered time wise, shall therefore stand for memory configurations which are in correspondence with the past experience of the observer. These configurations can be thought of as punches in a paper tape, impressions on a magnetic reel, configurations of a relay switching circuit, or even configurations of brain cells. We only require that they be capable of the interpretation \lq\lq The observer has experienced the succession of events $A,B,\dots,C$." (We shall sometimes write dots in a memory sequence, $[. .. A,B, ... ,C]$, to indicate the possible presence of previous memories which are irrelevant to the case being considered.)}
\end{quote}

The parameters in non-italic rectangular brackets cannot be how the succession of events experienced by the observer system is actually recorded in the observer wavefunction.  If they were recorded only in this way, then RSQM would need a formal mechanism for manipulating indices during an observation, which it does not provide. 

Apparently, these parameters are only an informal notation to draw attention to the idea that the observer wavefunction itself internally encodes the information about the succession of events experienced by the observer system.  The fact that the parameters in HEA 9 were indices in HET and promoted to parameters in HEA strongly supports this interpretation.  

Clearly, a universal wavefunction that accounts for every atom in a macroscopic observer apparatus encodes the information about any object system in a fully microscopic description of the macroscopic observer system.  But such a wavefunction is impossibly complex and the microscopic encoding of the information is extremely inefficient.  As already discussed here, there are many intermediate links in the chain of interactions that detect and transmit properties of a microscopic object system to a macroscopic recorder system in the observer system.  HET ignores these links at this point and continues by presenting the first definition of a good measurement.  

\subsection{RSQM definition of a good observation}  

The definition of a good observation of a property of a system in an eigenstate is given on pages 65 and 66 of HET and in almost identical form on page 458 of HEA):
\begin{quote}
\textit{A good observation of a quantity $A$ for a system $S$, with eigenfunctions $\phi_i$ for a system $S$, by an observer whose initial state is $\psi^O$, consists of an interaction which, in a specified period of time, transforms each (total) state
\begin{equation}
\tag{HEA 10} 
\psi^{S+O} = \phi_i \psi^O[\dots]
\end{equation}
 into a new state 
\begin{equation}
\tag{HEA 11} \label{S+O'}
\psi^{S+O'} = \phi_i \psi^O[\dots\alpha_i].
\end{equation}
where $\alpha_i$, characterizes\footnote{HEA footnote 7: \textit{It should be understood that $\psi^O[\dots\alpha_i]$ is a different state for each $i$. A more precise notation would write $\psi^O_i[\dots\alpha_i]$, but no confusion can arise if we simply let the $\psi^O_i$ be indexed only by the index of the memory configuration symbol.}} the state $\phi_i$. That is, we require that the system state, if it is an eigenstate, shall be unchanged, and (2) that the observer state shall change so as to describe an observer that is \lq\lq aware" of which eigenfunction it is; that is, some property is recorded in the memory of the observer which characterizes $\phi_i$, such as the eigenvalue. The requirement that the eigenstates for the system be unchanged is necessary if the observation is to be significant (repeatable), and the requirement that the 
observer state change in a manner which is different for each eigenfunction is necessary if we are to be able to call the interaction an observation at all.}
\end{quote} 

This definition does not explicitly state what happens to the object system during the \lq\lq specified period of time," much less the details of how the wavefunction evolves during this time.  The statement, \lq\lq That is, we require that the system state, if it is an eigenstate, shall be unchanged, \dots" can be interpreted in two different ways. 

It might mean that the definition of a good measurement imposes no restrictions on what happens to $\phi_i$ \lq\lq during the specific period of time during which $\psi^O[\dots]$ is transformed into $\psi^O[\dots\alpha_i]$ while requiring that $\phi_i$ represent the same state at the beginning and end of that period of time.  Alternatively, this definition could require that $\phi_i$ represent the same state during the entire observation period.  

Clearly, HE means the latter.  The former would require HET to abandon the development, definition, and flawed example of an RSQM measurement, which is a major part of the thesis. Still, it is perplexing that the RSQM definition of a good observation does not even mention the RSQM definition of a measurement. 

At this point, RSQM is a quantum mechanical theory that cannot be proven wrong because it is based on definitions that limit its scope to a negligible fraction of interactions.  HE next extrapolates conclusions drawn from the definition of a good observation to all interactions.  No justification for this extrapolation is given.   
      
\subsection{The RSQM theory of nothing becomes the RSQM theory of everything}

Following the definition of a good observation and a short discussion, HE imposes a restriction on the final state of an general object system in a general initial state. 
\begin{quote}
HET 66 and HEA 458: \textit{From the definition of a good observation we first
deduce the result of an observation upon a system which is \textbf{not} in an eigenstate of the observation. We know from our definition that the interaction transforms states
$\phi_i \psi^O[\dots]$ into states $\phi_i \psi^O[\dots\alpha_i]$. Consequently these solutions of the wave equation can be superposed to give the final state for the case of an arbitrary initial [object] system state. Thus if the initial [object] system state is not an eigenstate, but a general state $\sum_i a_i \phi_i$, the final total state will have the form:
\begin{equation}
\tag{HEA 12} 
\psi^{S+O'} = \sum_i a_i \phi_i \psi^O[\dots \alpha_i].
\end{equation}
This superposition principle continues to apply in the presence of further systems which do not interact during the measurement.}
\end{quote} 

HE explains the meaning of Eq. HEA 12 on page 459 of HEA :
\begin{quote}
\textit{We thus arrive at the following \textbf{picture}: $\dots$  Throughout all of a sequence of observation processes there is only one physical system representing the \textbf{observer}, yet there is no single unique state of the observer (which follows from the representations of interacting systems). Nevertheless, there is a representation in terms of a superposition, each element of which contains a definite observer state and a corresponding system state.  Thus with each succeeding observation (or interaction), the observer state \lq\lq branches" into a number of different states.  Each branch represents a different outcome of the measurement and the corresponding eigenstate for the object-system state.  All branches exist simultaneously in the superposition after any given sequence of observations. \newline
\mbox{\;\;\;\;} The \lq\lq trajectory" of the memory configuration of an observer performing a sequence of measurements is thus not a linear sequence of memory configurations, but a branching tree, with all possible outcomes existing simultaneously in a final superposition with various coefficients in the mathematical model. In any familiar memory device the branching does not continue indefinitely, but must stop at a point limited by the capacity of the memory.}
\end{quote}

HE deduced the form of Eq. HEA 12 from the definition of a good observation, which requires that $\phi_i$ remain unchanged during the entire duration of a good observation: 
\begin{quote}
\textit{The requirement that the eigenstates for the system be unchanged is necessary if the observation is to be significant (repeatable), and the requirement that the observer state change in a manner which is different for each eigenfunction is necessary if we are to be able to call the interaction an observation at all.}
\end{quote}
Yet HE extrapolated Eq. HEA 12 from this very limited set of object systems to the final state of any object system with any arbitrary initial-state wavefunction $\sum_i a_i \phi_i$.  

When treated as a deduction based on the definition of a good observation, the scope of Eq. HEA 12 is drastically reduced.  It can't describe nuclear fusion, fission, or the spontaneous decay of radioactive isotopes because not only is the final state different from the initial state, but the final system is different from the initial system. The same is true for chemical reactions.  Even catalysts change state during a reaction before they revert to their initial state. Furthermore, Eq. HEA 12 can't even describe electronic transitions that were the inspiration for the development of quantum mechanics.  

When treated as the final state for the case of an arbitrary initial [object] system state, Eq. HEA 12 is just plain wrong and RSQM is demoted from a theory of the behavior of a very small if not empty set of physical systems to a blatantly false theory of the behavior of all physical systems. The requirement that the state of an object system remain unchanged during the entire duration of an observation is the proximate cause of this problem.  

The concepts of same, different, and change are the foundation of human experience.  Experimental science is an attempt to qualify and quantify change, the former typically referred to as an observation (but not by Everett), the latter as a measurement. 

HET starts out by claiming that \lq\lq there are two fundamentally different ways in which the state function can change", and then postulates that the ideal (good)  observation of a value of a property of an eigenstate by an observer system requires that the eigenstate that describes value remain unchanged until the observer system records the value.  This contradicts the entire history of the observation and measurement of change. 

For instance, Newton's third law is generally translated into English as, \lq\lq For every action (force) in nature there is an equal and opposite reaction." The description of entanglement in CQM preserves this idea. 

It is true, that the mathematical formulation of quantum mechanics provides Process 2 to calculate the eigenvalue of a system in an eigenstate without changing the state of the system: $H \psi = \lambda_n \psi$.  Schrodinger's equation was designed to make this work.  

Unfortunately, there is no one-to-one mapping of Process 2 onto an experimental apparatus and a sufficiently detailed examination of any practical experiment in terms of qualitative observation or quantitative measurement of a property of an object system by an observer system will show that the state of the object system changes at least once during the experiment.   

\section{RSQM's last gasp}

Everett embedded the idea throughout HET that the interpretation of an experiment was entirely subjective, and that when experiments were interpreted without the subjective bias of the concepts of CQM, their  results would agree with the predictions of RSQM:
\begin{quote}\textit{
\begin{enumerate} 
  \item Furthermore, we shall deduce the probabilistic assertions of Process 1 as \textbf{subjective} appearances to such observers, thus placing the theory in correspondence with experience. (HET page 9) 
  \item In this sense the usual assertions of Process 1 appear to hold on a \textbf{subjective} level to each observer described by an element of the superposition. (HET page 10) 
  \item The validity of Process 1 as a \textbf{subjective} phenomenon is deduced, as well as the consistency of allowing several observers to interact with one another. (HET page 12) 
 \item We are faced with the task of making deductions about the appearance of phenomena on a \textbf{subjective level}, to observers which are considered as purely physical systems and are treated within the theory. In order to accomplish this it is necessary to identify some objective properties of such an observer with \textbf{subjective} knowledge (i.e., perceptions). (HET page 63)
 \item In order to make deductions about the \textbf{subjective} experience of an observer it is sufficient to examine the contents of the memory. (HET 64) 
 \item Our problem is, then, to treat the interaction of such observer-systems with other physical systems (observations), within the framework of wave mechanics, and to deduce the resulting memory configurations, which we can then interpret as the \textbf{subjective} experiences of the observers. (HET page 65) 
 \item In the language of \textbf{subjective} experience, the observer which is described by a typical element, $\psi'_{ij... k}$ of the superposition has perceived an apparently random sequence of definite results for the observations. HET page 70) 
 \item We have thus seen how pure wave mechanics, without any initial probability assertions, can lead to these notions on a \textbf{subjective} level, as appearances to observers. (HET page 78) 
 \item Our abstract discussion of observation is therefore logically complete, in the sense that our results for the \textbf{subjective} experience of observers are correct, if there are any observers at all describable by wave mechanics. (HET page 85) 
 \item Our theory of pure wave mechanics, to which we now return, must give equivalent results on the \textbf{subjective} level, since it leads to Process 1 there. (HET page 97) 
 \item The irreversibility of the measuring process is therefore, within our framework, simply a \textbf{subjective} manifestation reflecting the fact that in observation processes the state of the observer is transformed into a superposition of observer states, each element of which describes an observer who is irrevocably cut off from the remaining elements. (HET page 98) 
 \item We saw that the probabilistic assertions of the usual interpretation of quantum mechanics can be deduced from this theory, in a manner analogous to the methods of classical statistical mechanics, as \textbf{subjective} appearances to observers - ... (HET page 109) 
 \item Our theory in a certain sense bridges the positions of Einstein and Bohr, since the complete theory is quite objective and deterministic ("God does not play dice with the universe"), and yet on the \textbf{subjective} level, of assertions relative to observer states, it is probabilistic in the strong sense that there is no way for observers to make any predictions better than the limitations imposed by the uncertainty principle. (HET page 117) 
 \item Nevertheless, within the context of this theory, which is objectively deterministic, it develops that the probabilistic aspects of Process 1 reappear at the \textbf{subjective} level, as relative phenomena to observers. (HET page 117)
\end{enumerate}}
\end{quote}
Interestingly, \lq\lq subjective" only appears once in HEA and not exactly in this context. Perhaps Everett came to understand that this argument did not work.  

By the way, despite the assertion on page 85 of HET, \lq\lq Our abstract discussion of observation is therefore logically complete, in the sense that our results for the \textbf{subjective} experience of observers are correct, if there are any observers at all describable by wave mechanics." is not true.  What HE actually did demonstrate is the much weaker result: If the ratios of the weights $a^*a_i$ from Eq. 12 are identical to the ratios determined experimentally for a superposition in Eq. 12, then the RSQM "results for the \textbf{subjective} experience of observers are correct." 

However, the weights $a_i^*a_i$ in Eq. 12 fail spectacularly to describe the experience of observers in the case of spontaneous decay of excited sates of nuclei and atoms.  First, consider a good observation of a property of a system in an eigenstate, which is just a subset of cases described by Eq. 12.  Let the system be the three-baryon nuclear complex consisting of the tritium nucleus $\big(^3_1H^+\big)$ and the helium-3 nucleus $\big(^3_2H^{++}\big)$.  In this case,  the initial state of a tritium nucleus at time $t=0$ following the creation of a sample of tritium nuclei by any of a number of nuclear processes  can be written as
\begin{equation}
\psi^{S+O'}(0) = a(0)\phi(0)\psi^O(0, [\dots]),
\end{equation} 
where $a(0) = 1$, $\phi(0)$ is the wavefunction of the tritium nucleus and $\psi^O(0, [\dots])$ is the wavefunction of an energetic particle detector such as a Geiger-Mueller counter that has negligible influence on the decay of the tritium nuclei as is easily demonstrated experimentally.  There there is no $\alpha$ following the bracketed dots because the observation has not yet started. This expression is consistent with CQM as well as RSQM, but the factor  $\psi^O(t, [\dots])$  is superfluous in CQM because the observer system evolves independently of the three-baryon object system after formation of the excited state.  This is the usually the case in measurements of the properties of radioactive isotopes.  

According  to CQM, the final state of this total system at any specified time $t$ is well approximated by 
\begin{equation}
\psi^{S+O'}(t) = a(t)\phi_T(t) + \big(1-a^*(t)a(t)\big)^{1/2}\phi_{He}(t),
\end{equation}
where $a(t) = \exp(-t/2\tau)$, $\tau$ is the lifetimes of the tritium nucleus, $\phi_T(t)$  is the tritium nuclear wavefunction, and $\phi_{He}(t)$  is the helium-3 nuclear wavefunction.  

When first developed, Eq. 5 was thought by some to describe the state of each 3-baryon nucleus in the sample as a function of time.  In fact, it actually is the wavefunction that describes the mean evolution of the 3-baryon field as a function of time normalized to the original occupation number of the three-baryon field and the actual transition from $\phi_T(t)$ to $\phi_{He}(t)$ occurs in less than $10^{-30}\tau$. 

How do we know this?  One way is by observing individual transitions in different ways at different times relative to the creation of tritium samples\cite{Jones, Unterweger}.  Another way is by observing the increase of the final state with time\cite{Akulov3}.   The is another less direct way that may be more interesting to science fiction.  If the state of \textbf{each particle} in the original sample of tritium actually did evolve according to Eq. 5, then at time $t = \tau_{1/2} = \tau\ln(2)$ we could measure the physical and chemical properties of an entirely new element (Trelium-1/2) with wavefunction 
\begin{equation}
\psi^{S+O'}(\tau_{1/2}) = \sqrt{2}\Big[a(t)\phi_T(\tau_{1/2}) + \big(1-a^*(\tau_{1/2})a(\tau_{1/2})\big)^{1/2}\phi_{He}(\tau_{1/2})\Big].   
\end{equation}
These properties would differ substantially from those of both tritium and helium-3. In fact, the periodic table of isotopes would have a continuous fourth dimension for elements like trelium-$\beta$ for $0 < \beta < 1$.  The fact that no elements like trelium-$\beta$ have ever been reported does not mean that they cannot exist with some lifetime, it just means 1) that superposition with the ground state are not a byproduct of exciting a system to a more energetic state than the ground state if the energy difference between the states is large and that while Eq. 5 does describe the average time evolution of systems in a pure eigenstate, it does not describe the time evolution of the individual particles.  
 
 The fact that Eq. 5 describes fields rather than particles will be very important when more than one eigenfunction is included in Eq. 12.  But first the RSQM equivalent of Eq. 5 has to be compared with Eq. 5 of CQM.  

 According to RSQM, the final state of the initial state of the total system defined in Eq. 4 at time $t$ is given by 
\begin{equation}
\psi^{S+O'}(t) = a(t)\phi_T(t)\psi^O(t, [\dots,\alpha_T]),
\end{equation}  
where $a(t) = 1$ and $\alpha_T$ is some property of $\phi_T(t)$.   In light of the contents of the previous sections of this report, it is not surprising that there does not appear to be any way to interpret this result consistent with the facts of experimental physics  or Eq. 5 of CQM.   

The difference between Eq. 6 of CQM, which is consistent with the experimentally determined exponential decay rate, and Eq. 7, which is not, is the requirement in Eq. HEA 12 that $a(t) = 1$, which is imposed by the RSQM definition of a good observation. This requirement precludes spontaneous transitions of excited states to lower-energy excited states or to the ground state by conserving the single-eigenfunction state of the initial condition. 

This discussion does not demonstrate that superpositions of excited states and ground states are not possible. Instead, it only demonstrates that exciting a system to an excited state does not automatically create a superposition of the excited state and a ground state.  In fact, it is very difficult if not impossible to create a superposition of an excited state and a ground state with nominally monochromatic radiation with an energy bandwidth substantially smaller than the energy difference between the excited state and the ground state. 

Having established this fact, it is now time to apply Eq. HEA 12 to a system in a superposition of two eigenstates. For this exqample, consider a sample of $N$ isolated sodium atoms such as in a sodium vapor.  Let each of these atoms be in a 50/50 superposition of two sodium D-line states described by $\phi_1(t)$ and $\phi_2(t)$. 

The initial total state of each atom can be described by  
\begin{equation}
\psi^{S+O'}(0) = \big(a_1(0)\phi_1(0) + a_2(0)\phi_2(0)\big) \psi^O(0, [\dots]),
\end{equation} 
at the time $t=0$ from the formation the superposed excited states, where $a_1(0) = a_2(0) = \sqrt{0.5}$.  There there is no $\alpha_1$ or $\alpha_2$ following the dots because the observation has not yet started.  This expression is consistent with both CQM and RSQM, but the $\psi^O(t, [\dots])$ factor is superfluous in CQM if the observer system evolves independently of the object system after formation of the excited state, which is the usually the case in measurements of electronic transitions in atoms, ions, and molecules.  

According  to CQM, the final state of this total system at a specified time $t$ is well approximated by 
\begin{equation}
\psi^{S+O'}(t) = \bigg(a_1(t)\phi_1(t) +  a_2(t)\phi_2(t) +\sqrt{2-a_1(t)-a_2(t)}\phi_0(t)\bigg) /\sqrt{2},
\end{equation}
where $a_i(t) = \exp(-t/2\tau_i)$, $\tau_i$ is the lifetimes of the excited state described by $\phi_i(t)$, and the $\psi^O(t, [\dots])$  factor is superfluous if the observer system evolves independently of the object system after formation of the excited state, which is the usual case in measurements of the lifetime of excited states. 

In CQM, the time dependent expectation value of $\psi^{S+O'}(t)$ describes a fluctuating quantum beat superimposed on exponentially decaying populations of the two excited states.  Both the exponential decay and the quantum beats can be observed\cite{Carlsson} by measuring a fraction of the spontaneously emitted photons as a function of time.  

According to RSQM, the final state after the specified time $t$ is given by 
\begin{equation}
\psi^{S+O'}(t)  = a_1(t)\phi_1(t)\psi_1^O(t) [\dots, \alpha_1] + a_2(t)\phi_2(t) \psi_2^O(t, [\dots,\alpha_2])  
\end{equation} 
where $a_1(t) = a_2(t) = 1$ and the subscript of the final states of the observer system that HET suppresses in Eqs. HET 10 through HET 12 (as described in footnote 7 on page 17) are included here for clarity. No matter how much I try, I cannot explain how RSQM interprets this equation except to refer the reader to the original publications.  I think this is what Schlosshauer\cite{Schlosshauer} was referring to when he wrote, \lq\lq Everett never clearly spelled out how his theory was supposed to work." 

Needless to say, RSQM Eq. 10 is incompatible with CQM Eq. 9, which describes lifetime measurements of singly ionized alkali atoms quite well, and more complex or more highly ionized heavy ions less well, but still well enough to raise no questions about the scope of CQM.  

On the other hand, the differences between Eqs. 6 and 7 and between Eqs. 9 and 10 are not small differences of this nature.  These differences are qualitative differences that go well beyond quantitative differences in predicted lifetimes. The final states as a function of time are different and the difference increases with increasing time until the end of the specified time period during which the interaction is observed. Yet HE claims that these difference are all explained by the subjective nature of our experience. However, there is nothing subjective about changes of state.  

A description of the results of an experiment in terms of the subjective concepts of CQM theory is much more concise than a description from which all of the subjective concepts have been removed.  However, changes of state can be described entirely in terms of experimental methods to distinguish one state from another and experimental methods to separate one system from another without any recourse to theory. These are fundamental observations that babies learn without any theoretical concepts at all.  
No one needs to know any chemistry or physics, much less the theoretical ideas of quantum mechanics to distinguish a test tube full of clear water from a test tube with a precipitate in clear water.  Any theory that claims these are the same state of the same system contradicts the theory-free facts of our experience and is unable to make meaningful predictions about that experience. This is the fate of the RSQM due to the definitions of a good observation.   

\subsection{Synopsis}

The HET program to reformulate CQM started with a review of information theory tailored for treating measurements and observations in terms of correlations. This was followed by a review of CQM minus all material associated with measurement and observations.  This review was tailored for treating these topics in terms of correlations to replace the material removed from CQM.  

Following this review, HET then directly addressed quantum mechanical measurements in terms of correlations. Unlike CQM, which attempts to treat all interactions including measurements with the same formulation, HET developed its treatment of measurements in terms of interactions between a object system and an observer system that each satisfy very stringent requirements.   

Therefore, the formulation and conclusions drawn from this treatment apply only to the subset of all interactions that satisfy these requirements until these conclusions are derived or verified in some other more general way. This section also included a flawed example of an RSQM measurement that was intended to illustrate the generality and power of the RSQM definition of a measurement. 

However, that generality is an illusion. There is an error in the mathematical example. When corrected, the example applies only to constant (that is to say, zero energy) eigenfunctions and distributions).  In other words, the example actually illustrates how fatally restrictive the RSQM definition of a measurement actually is.  

Apparently unaware of this problem, HET proceeded to develop an RSQM theory of observation in terms of interactions that satisfy the same very stringent requirements imposed on interactions that qualify as RSQM measurements.  The main difference between the RSQM definition of a measurement and the RSQM definition of a good observation is that the former iss given in the limit as time increases without bound, whereas the latter is given in terms of a specific period of time. Another difference is that the latter imposes even more restrictions on the observer system than the former does on a measurement.  

HET then deduced the final state of the interaction of an observer-system wavefunction and a superposition of eigenfunctions that individually meet the requirements of a good observation of a property of the superposition. At the onset of the interaction, the observer system is described by a standard CQM tensor product of a wavefunction and a superposition.  

At the end of the interaction, the final state describes a superposition of tensor products.  Each term in the final state is the tensor product of an eigenfunction term in the original superposition and a different version of the observer wavefunction. Furthermore, each version of the observer wavefunction records the eigenvalue of the associated eigenfunction term.  However, there is no quantum beat between the individual terms in the superposition, yet all of these observer wavefunctions are just different states of a single observer.    

This would be of little interest since it applies only to superpositions of object functions that satisfy the restrictive definition of an RSQM good observation. However, in deducing the form of the wavefunction that describes the final state of this system, HET extrapolated it to describe the final state of any \lq\lq arbitrary initial system state" without any explicit motivation or deduction.  

It is possible that the motivation for this extrapolation was the flawed example described earlier since that example appears to be very general.  However, that generality is an illusion. There is an error in the mathematical example. When corrected, the example applies only to constant (that is to say, zero energy) eigenfunctions and distributions).  In other words, the example actually illustrates how preposterous the RSQM definitions of a measurement and of a good observation actually are.

\end{document}